\documentclass[11pt]{article}

\usepackage{graphicx}
\usepackage{authblk}
\usepackage{amsmath,amssymb}
\usepackage{fullpage}

\newtheorem{theorem}{Theorem}[section]
\newtheorem{lemma}[theorem]{Lemma}

\newcommand{\pair}[2] {\langle #1,#2\rangle}
\newcommand{\NS} {n_s}
\newcommand{\ND} {n_d}
\newcommand{\Hf} {h}
\newcommand{\D} {{\rm D}}
\newcommand{\LL} {{\rm LL}}
\newcommand{\LR} {{\rm LR}}
\newcommand{\GBs} {{\rm ~GB/s}}
\newcommand{\GBsdir} {{\rm ~GB/s/direction}}
\newcommand{\sigmap}[2] {\sigma^{(#1)}_{#2}}

\newcommand{\eat}[1] {}
\newcommand{\qed} {\hfill$\Box$}

\title{Mapping Strategies for the PERCS Architecture}

\author[1]{Venkatesan T. Chakaravarthy}
\author[2]{Naga Praveen Kumar Katta}
\author[1]{Monu Kedia}
\author[3]{Ramakrishnan Rajamony}
\author[4]{Aruna Ramanan}
\author[1]{Yogish Sabharwal}
\affil{IBM Research - India\\
\{vechakra,mokedia1,ysabharwal\}@in.ibm.com
}
\affil[2]{Princeton University, USA\\
nkatta@cs.princeton.edu
}
\affil[3]{
IBM Research - Austin, USA
rajamony@us.ibm.com
}
\affil[4]{
IBM USA
arunar@us.ibm.com
}

\begin{document}
\maketitle

\begin{abstract}
The PERCS system was designed by IBM in response to a DARPA challenge that called for a high-productivity
high-performance computing system. 
The IBM PERCS architecture is a two level direct network having low diameter and high bisection bandwidth.
Mapping and routing strategies play an important role in the performance of applications on such a topology.
In this paper, we study mapping strategies for PERCS architecture, 
that examine how to map tasks of a given job on to the physical processing nodes.
We develop and present fundamental principles for designing good mapping strategies that minimize congestion. 
This is achieved via a theoretical study of some common communication patterns under
both direct and indirect routing mechanisms supported by the architecture.
\end{abstract}

\section{Introduction}
\label{intro}
The PERCS supercomputer is the latest in a line of high performance
computing systems from IBM~\cite{rajamony2010ibmjrd}. Previously known as 
``POWER7-IH''~\cite{rajamony2010ibmjrd} and commercialized as the
Power 775~\cite{ibm-p775ga}, this system can
incorporate up to 64K Power7 processors with a high-radix,
high-bisection-bandwidth interconnect based on the IBM Hub
Chip~\cite{arimilli2010hoti, arimilli2010hotchips}. The system is a direct outcome of the  
similarly named PERCS project (Productive Easy-to-use Reliable Computing System) that was 
initiated by DARPA in 2002 with the goals of providing state-of-the-art 
sequential performance as well as interconnect performance exceeding the
state-of-the-art by orders of magnitude. With its integrated interconnect and
storage, the PERCS system is expected to provide high sustained performance on
a broad class of HPC workloads.

An important aspect of the PERCS architecture is the two-level direct-connect topology
provided by a Hub chip~\cite{arimilli2010hotchips}. 
In this setup, Hub chips (or {\em nodes}) are grouped in the form of cliques 
(called {\em supernodes}) and these cliques are then inter-connected.
A total of 47 links out of each Hub chip create an 
interconnect with no more than three hops required to reach any other Hub. 
Furthermore, the denseness of this interconnect yields high bisection
bandwidths and makes the system suitable for communication intensive workloads.
The PERCS topology is noteworthy in that the multiple links can be setup between a pair of supernodes.

A fundamental issue in parallel systems is the mapping of tasks of a given job to the processors
based on the communication pattern of the job and the characteristics of the underlying topology.
The mapping determines the congestion on the network links and hence the performance of the application.
The mapping problem has been extensively studied with respect to various communication topologies 
such as the torus, hypercube and fat-tree interconnects (see, for example, \cite{ASK06,BS87,ERP88,FLN89}).
Prudent mapping not only results in improved communication bandwidth but
also less inter-job interference. 

To the best of our knowledge, the work of Bhatale et al\cite{SCpaper} is the only prior study of the 
mapping problem on the PERCS architecture. They consider a sample set of communication patterns (
2D 5-point stencil, 4D 9-point stencil, Multicast)
and developed heuristics to perform the mapping. The idea was to divide the given job into blocks
and partition the processors into blocks (where each block consists of neighboring processors); the job blocks
are then mapped onto the processor blocks in a random manner. Different heuristics were obtained by changing the
block size. They conducted an extensive experimental evaluation to compare these heuristics.

{\bf Our contributions.}
The main contribution of this paper is to initiate a systematic approach for designing good 
mapping strategies for the PERCS architecture. 
Our aim is to develop mapping strategies based on sound theoretical
underpinnings derived via a formal analysis of the system. 

A complete study of the mapping problem for the PERCS architecture 
would ideally cater to handling multiple jobs executing simultaneously on the system;
the jobs having potentially different predominant communication patterns.
However, this can be a challenging task given the fact that the architecture is very recent and there is lack
of principled prior study on this architecture.
Being an initial study, we restrict the scope of this paper as follows:
(i) we consider only a single system-wide job;
(ii) we only consider the load on the links (link congestion), ignoring other protocol overheads -- this aids in focusing 
on the critical architectural aspects; and
(iii) we consider only two (but very contrasting) communication patterns.
We note that Bhatale et al.\cite{SCpaper} also consider only a single system-wide job and only a 
selected set of communication patterns.
We believe that the concepts and theoretical principles developed here will guide in a complete study of this problem.

Though our study has the above-mentioned limitations, 
we consider a wide range of system configurations determined by the following parameters:
\begin{itemize}
\item The system size (number of supernodes)
\item The number of links connecting any pair of supernodes 
\item Two different routing schemes: (i) direct routing, 
where data is sent over the direct link(s) connecting the supernodes;
(ii) indirect routing, where an intermediate supernode is used as a bounce point for redirecting data 
in order to improve the load balance.
\end{itemize}
In comparison, we note that Bhatale et al.\cite{SCpaper} consider only two system sizes 
(64 supernodes and 304 supernodes) whereas we consider a wider range system sizes.

An important aspect of the PERCS architecture is that the number of links between pairs of supernodes can be varied.
This wealth of choice in system connectivity needs to be carefully considered based on the needs of the
workloads that will be executed on the system. For randomized workloads like
those exemplified by the Graph500~\cite{graph500} and RandomAccess~\cite{hpcc} benchmarks, high
connectivity
will be beneficial. For many HPC workloads, a lower level of
connectivity may suffice. For instance, large classes of HPC workloads have been
shown to have fairly low-radix inter-task partnering
patterns~\cite{kerbyson-patterns}. For such workloads, choosing a
topology with a lower number of links between supernode pairs may result in
considerable cost savings.
An important feature of our study is that the number of links between a pair of supernodes is taken as a parameter
and can be varied. In comparison, prior work allow only a single link between a pair of supernodes.

The two communication patterns that we study are:
\begin{itemize}
\item {\bf Halo (two-dimensional 5-point Stencil)~\cite{WRF}: }
	Tasks are arranged in the form of a 2-D grid and each task communicates
	with its north, east, south and west neighbors. 
\item {\bf Transpose~\cite{hpcc}: }
	Tasks are arranged in the form of a 2-D grid and each task communicates
	with all the tasks in the same row and column as that of the given task. 
\end{itemize}
Note that these sample patterns are contrasting in nature. 
While the communication pattern of Halo is sparse, that of transpose is a fairly dense.
Halo is indicative of the small-partner-count patterns that dominate 
HPC workloads. Transpose is indicative of the performance of workloads that use spectral methods
such as the FFT. We next highlight some of the important contributions of our study with respect to each of these patterns.

For the Halo pattern, Bhatale et al.\cite{SCpaper} devised a mapping heuristic based on 
random mapping of task blocks to processor blocks (as mentioned earlier). 
We devise a deterministic strategy to map the task blocks to the processor blocks based on the theoretical properties
of modulo-arithmetic, that removes the randomness and provides up to a factor 2 improvement in throughput under 
direct routing.

To the best of our knowledge, our work is the first to study the transpose pattern for the PERCS architecture.
We argue that direct routing offers better throughput than indirect routing for this pattern. This is in strong contrast to all the patterns
considered by Bhatale et al.\cite{SCpaper} for which they experimentally demonstrated that the indirect routing is superior.
This shows that there are communication patterns that benefit under direct routing and therefore a study of the 
mapping problem would be incomplete without considering direct routing.

We also present experimental evaluation of the various mapping schemes using a simulator for computing throughput.
Our experiments confirm that the mapping based on mod-arithmetic is superior to the random mapping schemes for the Halo
pattern under direct routing by a factor of up to two. The experiments also validate that direct routing is better than 
indirect routing for transpose while indirect routing is better than direct routing for Halo.

Apart from PERCS, the concepts developed in this paper are more broadly applicable 
to a larger class of topologies categorized as multi-level direct networks 
with all-to-all connections at each level.
The Dragonfly topology \cite{dragonfly} that was introduced in 2008 articulates the technological
reasons for using high-radix routers while lowering the number of global cables
that criss-cross the system. 

\section{PERCS Architecture}
\label{architecture}
Our goal in this paper is to find a ``good'' mapping of tasks to compute elements 
for a variety of communication patterns and routing schemes.
We lay the groundwork in this section by describing the PERCS interconnection topology.
While discussing the topology and routing scheme, we simplify certain details of the actual system 
in 
order to make the model more understandable. 

The basic unit of the PERCS network is a Quad Chip Module or {\em node} which consists of four Power7 
processors~\cite{sinharoy2011jrd}. By virtue of being located in a tight package, the four
Power7 chips are fully connected to each other by very high bandwidth links operating at
48 GB/s/direction. In general, intra-QCM communication is high enough to be considered
``free'' for the purposes of this study. 
Eight nodes are physically co-located in a {\em drawer}.
The nodes in a drawer are connected in the form of a clique using bidirectional copper {\it LL-links}
(i.e., each pair of nodes in a drawer is connected by a dedicated LL-link) that provide 
$21 \GBsdir$ of bandwidth.
For ease of exposition, we will assume that each node also has a self-loop LL-link resulting in
a node having eight LL-links connecting it to itself and the other nodes within the drawer.

Four drawers combine to form a {\em supernode} with 
each pair of nodes in different drawers connected using a dedicated bidirectional optical {\it LR-link} operating 
at $5\GBsdir$.
Every node has twenty-four LR links that connect it to each of the twenty-four nodes in the other three drawers.

\begin{figure*}
\begin{tabular}{ccc}
\includegraphics[width=2.3in]{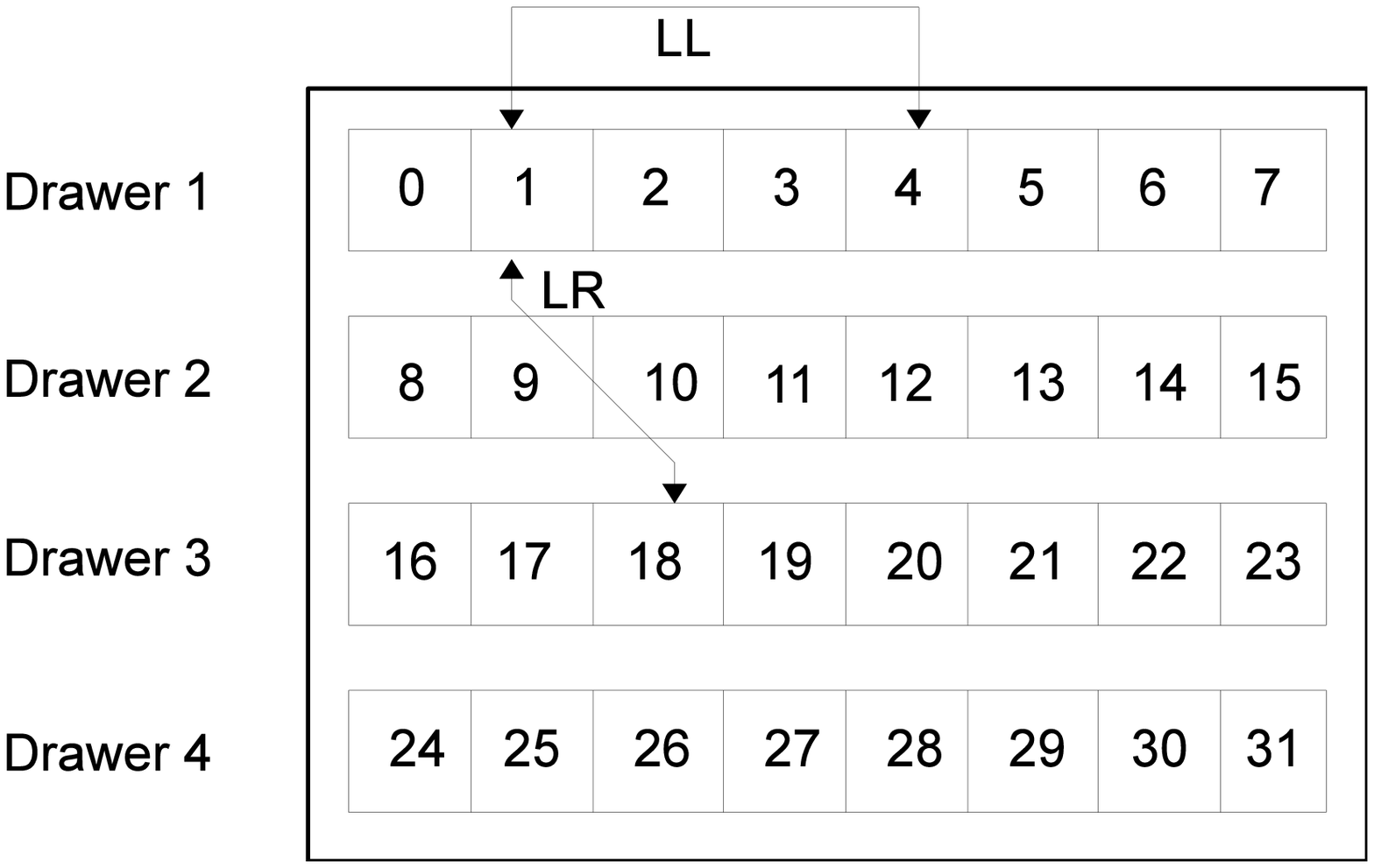}
& &
\includegraphics[width=4.2in]{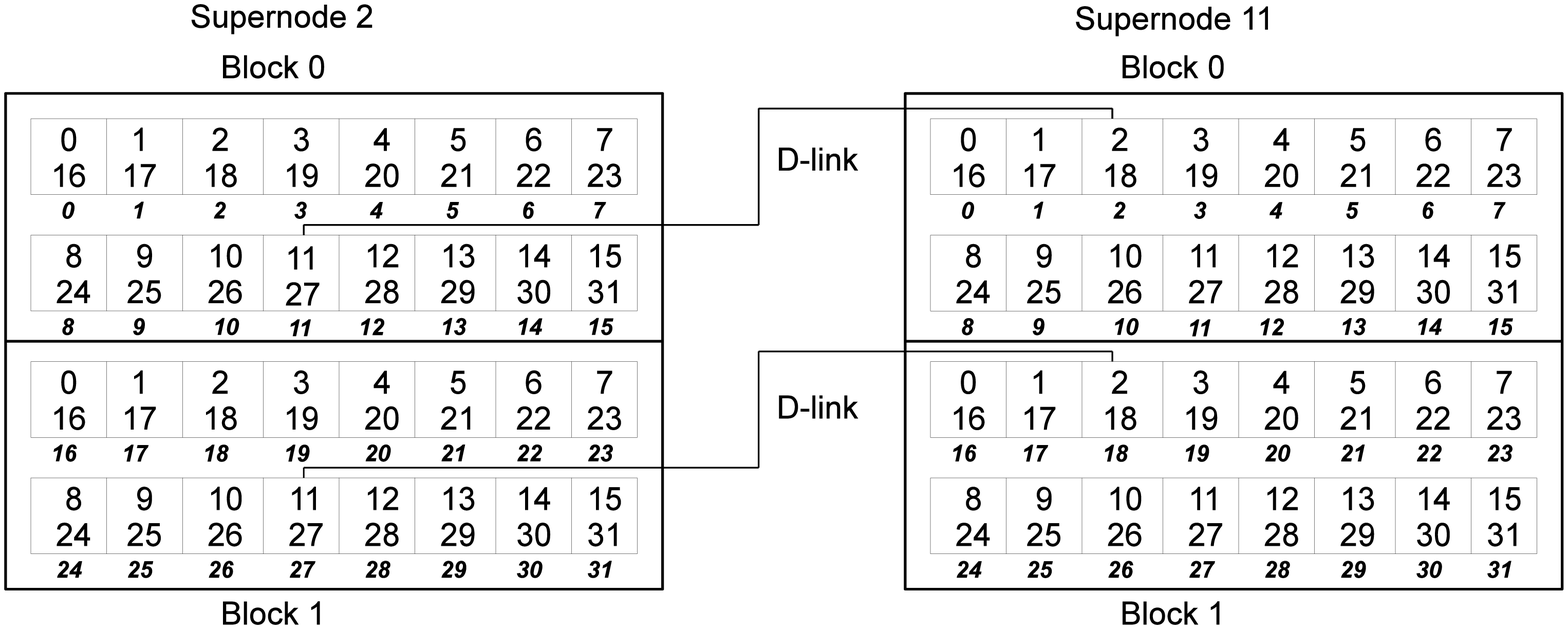}\\
(a) Supernode & & (b) D-link connectivity
\end{tabular}
\caption{Illustration of supernode and D-link connectivity}
\label{fig:AAA}
\end{figure*}

A supernode thus consists of a clique of thirty-two nodes connected by two type of links:
intra-drawer LL-links operating at $21 \GBsdir$ and inter-drawer LR-links operating at $5\GBsdir$.
In the rest of the paper, {\em we drop the designation ``/direction''} noting that all bandwidths
are treated as the bandwidth per direction.
Figure \ref{fig:AAA}(a) highlights one exemplar LL and LR link in a supernode.
We shall use the term L to denote both LL and LR-links.
Multiple supernodes are in turn connected via bidirectional {\em D-links}.
Each D link operates at $10 \GBs$ with the system design permitting multiple D-links between pairs of supernodes.

We shall use the term ${\NS}$ to denote the number of
supernodes in the system and the term ${\ND}$ to denote the number of D-links between pairs of supernode.
The tuple (${\NS}$, ${\ND}$) specifies a system. 
For instance, a $(32, 4)$ system consists of $32$ supernodes with each pair of supernodes being 
connected via four D-links.
The implementation requires that ${\ND}$ must divide $32$, the number of nodes in a supernode
yielding systems with six different ${\ND}$ values: $\{1,2,4,8,16,32\}$.

The value of $\ND$ determines the D-link wiring across supernodes.
Nodes within each supernode are divided into ${\ND}$ {\em buckets},
with each bucket having $32/{\ND}$ nodes. A D-link connects each bucket to the corresponding bucket in every 
other supernode -- therefore each bucket must have ${\NS}$ D-links originating from it connecting it to every
supernode in the system
(similar to our treatment of the LL links, we assume a self-loop D link for ease of exposition). Since a bucket
may have fewer nodes than are in a supernode, a node may be attached to multiple D links.
We call this the ${\ND}D$-topology and next describe a {\em template} that specifies where these $\NS$ D-links connect.
Let $W$ be the number of nodes in a bucket.
Then, the $j^{\rm th}$ bucket (where $0\leq j<\ND$) consists of nodes numbered $\{jW, 1+jW, \ldots, (W-1)+jW\}$.
Each bucket has $\NS$ D-links connecting it to every supernode in the system.
Within this bucket, the D-link connecting to supernode $b$ will originate from the node $v$ numbered $jW+(b\bmod W)$.
We refer to this as the supernode $b$ being {\em incident} on the node $v$. Thus, each supernode will be 
incident on $\ND$ nodes.

{\it Illustration: }Figure \ref{fig:AAA}(b) illustrates the D-link connectivity for a 32-supernode system
with two D-links between every pair of supernodes. This is an ($\NS=32$, $\ND=2$) system that uses 
a $2D$-topology. In this system, supernode $1$ is incident on nodes $\{1,17\}$.
Since $\ND=2$, there are two buckets each with sixteen nodes.
Bucket 0 consists of nodes in the first two drawers numbered $\{0,1,\ldots, 15\}$.
Bucket 1 consists of the remaining two drawers with nodes numbered $\{16,17,\ldots,31\}$.
Thirty-two D links go from each bucket to the thirty-two supernodes in the system.
Every supernode uses this template. Thus supernodes 0 and 16 are incident 
on node 0 in each bucket while supernodes 2 and 18 are incident on node 2 in each bucket.
The figure explicitly calls out the D-link connectivity between supernode 2 on the left and supernode 11 on the right.
Two D-links connect this pair of supernodes. In bucket 0, the D-link is between node 11 of supernode 2 
and node 2 of supernode 11. The same template is followed in bucket 1,
with a D-link between node 27 of supernode 2 (which is the same as node 11 in bucket 1) and node 18 of 
supernode 11 (which is the same as node 2 in bucket 1).

As a second example, consider a ($\NS=16$, $\ND=8$) system.
This system will have eight buckets per supernode each with four nodes.
Bucket 0 will contain the four nodes in the first half of the first drawer: $\{0,1,2,3\}$ 
while bucket 7 will consist of the nodes in the second half of the last drawer: $\{28,29,30,31\}$ 

Any node is indexed by a tuple $\pair{a}{u}$, where $a$ is the supernode number
($0\leq a\leq \NS-1$) and $u$ is the node number ($0\leq u\leq 31$).
For instance, $\pair{11}{13}$ will be node $13$ in supernode $11$.

We next define the notion of {\em D-port utility factor}, which is useful in our analysis.
The {\em D-port utility factor} $\Hf$ is defined to be the average number of D-links originating
from each node; formally, we define $\Hf = \NS/W$. 
Equivalently $\Hf=\NS\ND/32$.
For expositional ease, we only consider systems where $\Hf$ is an integer in this paper.
For instance, the $32$-supernode system of Figure \ref{fig:AAA}(b) 
has $h=2$ (i.e., two D-links originate and terminate at every node).
This terminology arises from the fact that nodes are equipped with optical transceiver ports that accept the D-links  
and $\Hf$ counts the number of D-ports used per node.

The Power 775 implements a maximum of $16$ D-ports at every node implying an $\Hf$ factor of at most $16$.
This means that $\NS\ND$ must be at most 512 capping the maximum system size at 512 supernodes.


The design supports two routing methods for inter-supernode communication.
In the direct routing scheme, the D-links directly connecting these supernodes are used to transfer data.
In the indirect routing scheme, an intermediate supernode is used for redirecting data
to the destination supernode. The direct routing scheme employs one D traversal,
while the indirect routing scheme uses two D traversal.
The advantage with indirect routing is in its ability to offer better load 
balancing~\cite{arimilli2010hoti,rajamony2010ibmjrd}. 
A formal description of these routing schemes will be provided in subsequent sections.

\section{Problem Statement}
\label{problem}
We now define the mapping problem precisely and introduce different communication patterns 
studied in the paper.

Every supernode has 32 nodes, each containing four Power7 processors.
We assume that each processor can execute a compute task yielding 
$128 \times {\NS}$ tasks system-wide.
The communication employed by the tasks in a job constitutes a {\em communication pattern} 
that can be visualized as a graph where the vertices denote tasks 
and directed edges denote inter-task communication.
The problem we investigate here is to map tasks to processors
so as to maximize system throughput taking both topology and routing into
consideration.

Prior work has observed that a small number of low-radix partnering patterns dominate most
HPC workloads~\cite{kerbyson-patterns}. Our analysis focuses on the following two patterns:
Halo and Transpose; it shows how task mapping can be improved for high-radix interconnects such as that found in 
PERCS.

{\bf Throughput Analysis: }
For our analysis, 
we assume that each task has a total of one unit of data 
to send to all of its neighbors. In the case of Halo,  the amount of data
sent to the neighbors is uniform. 
In the case of transpose, $(1/2)$ unit of is distributed uniformly
among the tasks in the same row and $(1/2)$ unit is distributed uniformly among the tasks
in the same column.  
In the case of Halo, each task $x$ will send $(1/4)$ units
of data to each of its four neighbors; in the case of transpose job consisting of $32$ rows and $64$ columns,
each task $x$ will send $(1/128)$ unit of data to each task on the same row
and $(1/64)$ unit of data to each task on the same column.

Consider a job $J$ and a function $\pi$ mapping the tasks of $J$ to the physical Power7 processors.
Data transfer between task $x$ and $y$ takes place between the physical processors
$u=\pi(x)$ and $v=\pi(y)$. 
The specific routing choice (direct or indirect) employed determines the set of paths over which the data is sent 
and thence the amount of load placed on every link in the paths.
We then obtain the {\em total load} $\ell(e)$ on a link $e$ by calculating the load contributions 
from every pair of communicating tasks that use that link.

Let $L_{LL}$ be the maximum of $\ell(e)$ over all the LL-links.
Similarly, let $L_{LR}$ and $L_{D}$ denote the maximum total load over all the LR and D-links.
Recall that the LL, LR and D links have different bandwidths of 21, 5, and 10 {\GBs} respectively.
These bandwidth differences arise from the very different cost, power, and capabilities that each
transport offers. Continuing our analysis, we normalize these numbers
by dividing $L_{LL}$ by 21, $L_{LR}$ by $5$ and $L_{D}$ by 10.
Let $t=\max\{L_{LL}/21, L_{LR}/5, L_{D}/10\}$. The value $t$ is a measure of congestion in the system
and a measure of the overall time it will take for the job to complete execution.
The value $1/t$ provides the throughput per task.
In other words, $(1/t)$ is maximum amount of data that every task can send
so that every LL, LR and D-link gets a total load of at most 21 {\GBs}, 5 {\GBs} and 10 {\GBs} respectively.
If each task sends $(1/t)$ GB of data, then the whole communication
will be completed in one second. 
Since four tasks execute on each node, the value $(4/t)$ provides the throughput per node.

We define the {\em throughput} of the mapping $\pi$ to be $\tau(\pi) = (4/t) \GBs$.
We also define throughput with respect each type of link:
(i) $\tau_{\LL}(\pi) = (1/L_{LL})\times 4\times 21$; 
(ii) $\tau_{\LR}(\pi) = (1/L_{LR})\times 4\times 5$; 
(iii) $\tau_{\D}(\pi) = (1/L_{D})\times 4\times 10$.
Notice that $\tau(\pi)$ is the minimum over the above three quantities.

{\bf Mapping Problem: }Given a specific job consisting of $n=128\NS$ tasks and a specific routing scheme, 
the problem is to devise a mapping $\pi$ having high throughput $\tau(\pi)$.

\section{Direct Routing}
\label{direct}
We first describe the direct routing scheme. Then, we study the mapping problem 
for the Halo and Transpose patterns.

\subsection{Routing Scheme}
\label{direct-routing}
We first describe intra-supernode routing, which is used for communication between two nodes
within the same supernode. Two types of routing schemes are supported for this scenario.

\begin{itemize}
\item
{\bf Single Hop: }All the communication between any two nodes is performed via the LL or LR link connecting them.
Figure \ref{fig:AAA}(a) shows
the single-hop route for communication between nodes 1 and 4,
as well as nodes 1 and 18.
\item
{\bf Striped: }Communication between nodes $u$ and $v$
is directed through one of the nodes in the same drawer as $u$ giving rise to LL-LL or LL-LR path.
A message from $u$ to $v$ is striped into multiple packets that are sent over the above eight paths.
If node $u$ has 1 unit of data to send, each path will receive $(1/8)$ units of data.
\end{itemize}

To illustrate striped routing, first consider the case of intra-drawer communication with 
node 0 sending data to node 1.
This transfer will be striped over eight paths of the form $0 \rightarrow x \rightarrow 1$, 
where any of the eight nodes ($\{0,1, \ldots, 7\}$)
present in the drawer can be used as the bounce point.
Note that we permit nodes 0 and 1 to also act as the bounce point;
in this case, we imagine that the LL self-loop at the node is utilized.
Next consider the case of inter-drawer communication with 
node 0 sending data to node 8. These nodes are in different drawers and we 
stripe it over eight paths of the form $0 \rightarrow x \rightarrow 8$, 
where the bounce point $x$ can be any of the eight nodes present in the first drawer ($\{0,1, \ldots, 7\}$).
If the amount of data is 1 unit, $1/8$ units of the sent will be sent over each path.
As before, node 0 can itself also act as a bounce point.
We thus stripe intra-drawer communication
over eight paths of the form LL-LL and inter-drawer communication
over eight paths of the form LL-LR.

Striped routing adds more load on the LL links compared to single-hop routing.  However,
it enables better link load balancing by using more of the higher bandwidth LL links.
In the rest of the paper, we only consider striped routing for intra-supernode
communications.

We next describe inter-supernode communication under direct routing when nodes $u$ and $v$
need to communicate but are located in different supernodes.
The destination supernode in incident on the source supernode through 
${\ND}$ D-links. Data transfer between the nodes is striped over these ${\ND}$ links
through paths of the form L $\rightarrow$ D $\rightarrow$ L.
If node $u$ has 1 unit of data to send, each path will receive $1/\ND$ units of data.

For an illustration, refer to Figure \ref{fig:AAA} (b) and consider node $1$ in supernode $2$ 
needing to send data to node $31$ in supernode $11$.
This communication will be striped over the two D-links shown in the figure utilizing the
following L-D-L paths:
(i) $\pair{2}{1} \rightarrow \pair{2}{11} \rightarrow \pair{11}{2} \rightarrow \pair{11}{31}$;
(ii) $\pair{2}{1} \rightarrow \pair{2}{27} \rightarrow \pair{11}{18} \rightarrow \pair{11}{31}$.

\subsection{Principles}
\label{principles}
In this section, we discuss some general principles that are useful 
in designing good mapping strategies for a general communication pattern $J$.

Developing a mapping strategy proceeds in two steps.
In the first step, we map each task to a supernode;
in the second step, we map the tasks to individual nodes within the supernodes.
Note that the D-links are used only for inter-supernode communication
causing the D-link throughput to be determined solely by the first mapping. D links are
the long global cables in the system.

Our experience suggests that it is often difficult to optimize on all the three types of throughputs.
A useful rule of thumb is to focus on D-link throughput while designing the first
mapping and consider LR-link and LL-link throughputs in the second step. Decomposing the
mapping problem in this manner makes it tractable.

Let us now consider the first step of mapping tasks to supernodes.
We view the mapping as a process of coloring of the vertices (i.e., tasks)
in the communication graph using the set of colors $\{0,1,2,\ldots,\NS-1\}$. 
We must ensure that all the colors appear an equal number of times 
($128$ times, since each supernode has 128 processors).
In order to obtain high D-link throughput, the coloring should ensure two properties: 
\begin{itemize}
\item
{\it Load Reduction: }A color class must form a dense subgraph with as few 
outgoing edges as possible. Doing so minimizes the total load contribution on the D-links.
\item
{\it Load Distribution: }For any pair of color classes, the number of edges between the color classes must be minimized.
This way the load on the bundle of $\ND$ D-links connecting any pair of supernodes will be minimized.
One way to achieve the above goal is to ensure that for each color class, 
the neighbors of the tasks in the color class are uniformly distributed across other color classes.
\end{itemize}

Let us now consider the second step of mapping tasks to individual nodes.
Consider an $L$-link going from a node $u$ to a node $v$.
For inter-supernode communication, link $e$ is used both when $u$ sends data
to any supernode incident on $v$ and when $v$ receives data from
any supernode incident on $u$. So, we must ensure that the eight tasks mapped
to nodes $u$ and $v$ do not have their neighbors concentrated on the above 
two types of supernodes. Doing so reduces the load imposed on $e$ due to inter-supernode
communication.

\subsection{Mapping Strategies for Halo}
\label{halo-direct}

In this section, we consider the Halo pattern. We first discuss certain natural strategies
and then present a scheme whose design is based on the properties of modulo arithmetic.
We shall mainly focus on the D-link throughput. However, we will ensure that LR-link and LL-link
throughput are not significantly compromised in the process.

Let number of processors $n$ is given by $n=128\NS$, since there are $128$ processors in a supernodes;
the number of tasks in the job is also $n$.
Consider the job as a $P\times Q$ grid consisting of $P$ rows and $Q$ columns such that $PQ=n=128\NS$.
Without loss of generality, assume that $P\leq Q$.
A simple strategy for mapping any pattern is to map the tasks sequentially
to the processors. This can be considered as a {\em default mapping}:
the $128\NS$ processors are indexed sequentially and the task having rank $j$
is mapped to the processor having index $j$.
Bhatale et al. have proposed certain natural strategies
that are better than default mapping based on the idea of creating 
blocks~\cite{SCpaper}.
We present an overview of these strategies.

\subsubsection*{Blocking Strategies}
Let $\alpha$ and $\beta$ be numbers such that they divide $P$ and $Q$, respectively.
The $P\times Q$ grid is divided into blocks of size $\alpha\times \beta$
resulting in $(P/\alpha) \times (Q/\beta)$ blocks.
Bhatale et. al. \cite{SCpaper} obtain different mapping schemes by choosing different 
values for $\alpha$ and $\beta$: 
\begin{itemize}
\item Node blocking: $\alpha\times \beta$ = $2\times 2$; fits in a node.
\item Drawer blocking: $\alpha\times \beta$ = $4\times 8$; fits in a drawer.
\item Supernode blocking: $\alpha\times \beta$ = $8\times 16$; fits in a supernode.
\end{itemize}
The next step is to map the tasks to processors which we accomplish
by dividing processors also into blocks:
in the case of drawer blocking, each drawer constitutes a processor block;
in the case of node and supernode blocking, each node and each supernode become a block, resp.
Processor and task blocks are indexed in a canonical manner.
Then the task blocks are mapped to processor blocks in one of the two ways:
\begin{itemize}
\item 
Sequential: A task block having index $j$ is mapped to processor block having index $j$.
\item
Random: Task blocks are mapped to processor blocks in a random manner.
\end{itemize}

Bhatale et al.\cite{SCpaper} proposed the above schemes and presented 
an experimental evaluation based on the BigSim simulator~\cite{bigsim}.
Below, we present an theoretical analysis of the throughput of these schemes.
Based on the ideas discussed here we develop an improved mapping strategy.

\subsubsection*{Blocking Strategies (An Informal Analysis)}
In the Halo pattern, a block is a dense subgraph and so, creating blocks reduces load on the different types of 
links.
We analyze the D-link throughput of the blocking schemes.
Consider an $\alpha\times \beta$ block. With respect to the task grid, 
the number of edges going out of the four sides this block is $2(\alpha+\beta)$.
Each of these edges carry $(1/4)$ units of data 
(since each task sends $(1/4)$ units of data to each of its four neighbors).
Thus, the amount of data going out of each block is $(\alpha+\beta)/2$.
Each supernode has $128/(\alpha\beta)$ blocks.
If no two adjacent task blocks are mapped to the same supernode,
the data leaving each supernode is 
\begin{eqnarray}
\label{eqn:CCC}
\frac{128}{\alpha\beta} \times \frac{\alpha+\beta}{2} = \frac{64(\alpha+\beta)}{\alpha\beta}.
\end{eqnarray}
For the node-blocking, drawer-blocking and supernode-blocking schemes, 
we see that the above quantity is $64$, $24$ and $12$ units, respectively.
Therefore, we note that supernode blocking is superior in reducing the overall load on the D links.
On the outset, it seems using larger blocks (as in supernode blocking) is
better than using smaller blocks (as in node blocking).
However, in determining the final throughput, it is important to consider the distribution of the 
load on the D links. This is discussed next.

Supernode blocking analysis is straightforward.
Recall that in this case $\alpha=8$ and $\beta=16$
Each supernode $a$ contains a single block whose four neighbors
are mapped on to four other distinct supernodes.
The data sent from $a$ to its northern and southern neighbors is $(\beta/4)=4$ units
and the data to sent to its eastern and western neighbors is $(\alpha/4)=2$ units.
The maximum amount of data sent from any supernode to any other supernode is $4$ units.
Since there are $\ND$ D links between any pair of supernodes, the maximum load
imposed on any D link is $4/\ND$ units. 
The throughput with respect to D links is therefore $(\ND/4)\times 4\times 10=10\ND$ \GBs.
Notice that the D link load in not well distributed, since from any supernode
data is sent only to the four neighboring supernodes causing 
the D links to the other supernodes to not be utilized.

In the above discussion, we managed to derive the exact D-link throughput for the supernode blocking scheme.
Performing a similar calculation for the expected D-link throughput of the 
node and drawer blocking schemes under random mappings is difficult.
However, we can make some qualitative remarks on these schemes.
We saw that compared to supernode blocking, 
the node and drawer blocking schemes put more load on the D links.
However, the latter schemes distribute the load more uniformly over the D links.
This is because the block sizes are smaller and so each supernode receives more number
of blocks. Consequently, for any supernode $a$, the number of blocks adjacent to the blocks
in $a$ is higher. As a result, under random mapping, these larger number of neighboring
blocks get more uniformly distributed over the other supernodes.
As an analogy, consider throwing balls (neighboring blocks)
randomly into bins (supernodes); the distribution gets more uniform as the number of balls
is increased.
Another important factor determining the overall throughput is the number of supernodes $\NS$.
As the number of supernodes increase, each supernode will receive lesser number of neighboring
blocks reducing the load on the D-links. 
Our experimental study 
compares 
the performance of different blocking schemes -- we show there 
that the blocking schemes achieve a D link throughput
between $10\ND$ and $20\ND$ \GBs, with the throughput increasing as the number of
supernodes is increased.

These observations motivate an improved mapping scheme called {\em mod color Mapping}.

\subsubsection*{Mod Color Mapping Scheme}
Similar to the blocking strategies discussed until now, the mod-coloring scheme is also based on dividing the input grid into blocks.
We first describe the process of mapping tasks to supernodes.
Let the input grid be of size $P\times Q$ where $PQ=n=128\NS$.
Consider any block size $\alpha\times\beta$ such that $\alpha$ and $\beta$ divide $P$ and $Q$, respectively.
Divide the input $P\times Q$ grid into blocks of size $\alpha\times \beta$.
Then $f=128/(\alpha\beta)$ blocks must be mapped to each supernode.
These blocks can be visualized as points in a smaller grid of size $p\times q$,
where $p=P/\alpha$ and $q=Q/\beta$.
Next, imagine each supernode to be a {\em color} and consider the set of
supernodes $\{0,1,\ldots, \NS - 1\}$  as a set of colors 
Then, carrying out a block-to-supernode mapping can be viewed as the process of 
coloring each point in the $p\times q$ grid with a color from the above set.
Each color appears $f$ times in the smaller grid and each color has four 
neighbors (east, west, north, south) in each appearance.
Each color has $4f$ neighbors over all appearances.
We say that a coloring scheme is {\em perfect}, 
iff for any color $c$, the $4f$ neighbors of $c$ are all distinct.
Figure \ref{fig:GGG} illustrated a perfect coloring for the case of 
$p=8$ and $q=8$; the number colors is $32$ and $f=2$; there are $32$ colors
each appearing exactly twice. The neighbors of 21 are $\{20,11,22,19,16,23,18,31\}$,
which are all distinct. A process of enumeration can show that the coloring is indeed perfect.

\begin{figure}
\begin{center}
\begin{footnotesize}
\begin{tabular}{cccccccc}
0 & 1 & 2 & 3 & 4 & 5 & 6 & 7\\
2 & 7 & 4 & 1 & 6 & 3 & 0 & 5\\
8 & 9 & 10 & 11 & 12 & 13 & 14 & 15\\
10 & 15 & 12 & 9 & 14 & 11 & 8 & 13\\
16 & 17 & 18 & 19 & 20 & 21 & 22 & 23\\
18 & 23 & 20 & 17 & 22 & 19 & 16 & 21\\
24 & 25 & 26 & 27 & 28 & 29 & 30 & 31\\
26 & 31 & 28 & 25 & 30 & 27 & 24 & 19
\end{tabular}
\end{footnotesize}
\end{center}
\caption{An example perfect coloring}
\label{fig:GGG}
\end{figure}

Assume now that we have constructed a perfect coloring scheme.
In terms of the blocks-to-supernode mapping $\pi$, this means that for any supernode $a$,
all the blocks in $a$ put together have at most one neighboring block
in any other supernode.  This means that the amount of data sent from 
from a supernode $a$ to any supernode $b$ is one of the following:
(i) no data is sent if they do not share a pair of neighboring blocks;
(ii) the data sent is $(\alpha/4)$ if the shared pair of blocks are neighbors in the east-west direction;
(iii) the data sent is $(\beta/4)$ if the shared pair of blocks are north-south neighbors.

Overall, the data sent from any supernode to any other supernode is at most $\max\{\alpha/4, \beta/4\}$.
It follows that the data sent on any D link is atmost $\max\{\alpha,\beta\}/(4\ND)$.
Therefore, for any perfect coloring scheme, the D-link throughput is:
\begin{eqnarray}
\tau_{\D}(\pi)&=&\frac{4\ND}{\max\{\alpha,\beta\}} \times 4\times 10.
\label{eqn:AAA}
\end{eqnarray}
One strategy for coloring the grid is to assign colors randomly
and this corresponds to the random block mapping strategies.

Equation~\ref{eqn:AAA} shows that the throughput obtained from a perfect coloring scheme
increases as the block size decreases. Thus, while the 
Bhatale-inspired supernode-blocking is a perfect coloring scheme,
its use of a large block size of $\alpha=8$ and $\beta=16$ 
results in a D-link throughput of $10\ND$ \GBs. 

In this context, our main technical result is a perfect coloring scheme 
with a block size of $\alpha=\beta=8$.
In this case, we see that $f=2$ (so that each supernode gets two blocks
or equivalently, each color appears exactly twice in the $p\times q$ grid).
The number of colors is $k=\NS=(pq)/2$.
The following lemma establishes the coloring scheme claimed above.

\begin{lemma}
\label{LEM:FFF}
Let $p$ be any multiple of four and $q\geq 8$ be any power of two
and the number of colors be $k=(pq)/2$.
Then there exists a perfect coloring scheme for the $p\times q$ grid. 
\end{lemma}

A proof sketch is provided below. A full proof is given in the Appendix \ref{app:mod-color}.
The proof uses a construction based on the properties of modulo arithmetic. 
For our choice of the block size $p$ is $P/8$ and $q$ is $Q/8$. In order to apply the lemma, 
we require that $P$ is a multiple of $32$ and $Q$ is a power of two at least $64$.
Suppose the grid satisfies the above properties.
Then Equation \ref{eqn:AAA} shows that we get a throughput of $20\ND$ \GBs.

{\it Remark: }Observe that if we could design a perfect coloring scheme with block size $4\times 4$,
the D-link throughput obtained will be $40\ND$ \GBs.
This as an interesting open problem.

Using the coloring scheme given by Lemma \ref{LEM:FFF}, we can map the $8\times 8$ blocks to
supernodes. We now focus on mapping the tasks within the blocks to individual nodes of the supernodes.
We divide the each block into $2\times 2$ {\em quads} and map the quads to the nodes of the supernode.
In order to guarantee a high throughput on the L-links, a careful mapping of quads to nodes 
is necessary. We leave this for future work.
In this paper, we adopt a simple strategy of mapping quads to nodes sequentially.
Our experimental analysis shows that the simple strategy works reasonably well.

To summarize, we presented the mod-color mapping scheme and proved that it offers a 
D-link throughput of $20\ND$ \GBs. We have also showed that the new scheme 
offers a $2\times$ improvement 
in D-link throughput over the supernode blocking scheme.
We later show that these observations persist in our experimental evaluations 
with the mod-coloring scheme outperforming other blocking schemes by as much as factor of
2. For a $P\times Q$, the mod-color scheme does require 
that $P$ be a multiple of $32$, and $Q$ be a power of two with $Q\geq 64$.
However, we believe that the framework developed here will be useful not only in
designing good strategies for general grid sizes, but also in obtaining even higher throughputs.

\subsubsection*{Proof Sketch of Lemma \ref{LEM:FFF}}
The proof uses the notion of {\em nice permutations}.
Let $\Omega$ be a set of $q$ elements. For a permutation $\sigma$ over $\Omega$
and $0\leq i<q$, let $\sigma(i)$ denote the symbol appearing in the $i$th position.
Consider two permutations $\sigma_1$ and $\sigma_2$ over $\Omega$.
Let us view the two permutations as $2\times q$ grid, where the first row is filled
with $\sigma_1$ and the second row is filled with $\sigma_2$.
Let $x$ be a symbol in $\Omega$. The copy of $x$ in the first row has 
three neighbors: left, right and down. Similarly, the copy of $x$ in the second
row has three neighbors: left, right and up. We say that the two permutations
are {\em nice}, if for any symbol $x$ all its six neighbors are distinct.

We claim that if $q\geq 8$ and $q$ is a power of two, then there always
a pair of nice permutations for $\Omega$. The proof of claim goes as follows.
We take $\sigma_1$ to be the identity permutation.
The permutation $\sigma_2$ is defined as follows: for $0\leq i<q$,
$\sigma_2(i)=(5i+2)\pmod{q}$. It can be shown that $\sigma_1$ and $\sigma_2$ form a nice pair of permutations.

Now let us prove Lemma \ref{LEM:FFF}.
We divide the set of $(pq/2)$ colors into $(p/2)$ groups of size $q/2$ each. 
Let these groups be $g_0, g_1, \ldots, g_{(p/2)-1}$
Then, for $0\leq i<(p/2)$, we color the points on the rows $2i$ and $2i+1$ as follows:
first we obtain a pair of nice permutations $\sigma_1$ and $\sigma_2$ for $g_i$; then we color the points
on the row $2i$ with the permutation $g_i$ and we color the points on the
row $2i+1$ with the permutation $\sigma_2$.
We can show that the coloring is perfect.

\subsection{Mapping Strategies for Transpose}
\label{transpose-direct}
In this section, we study the problem of designing good mapping strategies
for the Transpose pattern under direct routing.
We first focus on the D-links.

Let the input job be a grid of size $P\times Q$, where $P$ is the number of rows
and $Q$ is the number of columns, and $P\times Q=n=128\NS$ is the total number of tasks.
Any mapping strategy $\pi$ must divide the job in to $\NS$ groups each containing $128$ tasks,
and map a single group to each supernode.
In order to reduce the load on the D-links, it is important that the groups must have
high intra-group communication and low inter-group communication;
in other words, each group should be a dense subgraph in the job communication graph.
Recall that in the case of Transpose, each task sends $(1/(2Q)$ units of data to all
the tasks on its row and $(1/(2P))$ units of data to all the tasks on its column.
In terms of the communication graph, each row forms a clique and each column forms a clique.
Thus, there are two good mapping strategies: (i) row-wise mapping: map a set of rows
to each supernode; (ii) column-wise mapping: map a set of columns to each supernode.
Consider row-wise mapping
Assume that $Q\leq 128$ and that $Q$ is a power of two (so that
$Q$ divides $128$). Then, we map $128/Q$ rows to each supernode.
Let us compute the D-link throughput for the row-wise mapping. 
Let $e$ be a D-link going from a supernode $a$ to some other supernode $b$.
Any task $x$ sends $(1/(2P))$ units data to each task on its column (including itself).
The supernode $b$ contains $128/Q$ rows and from each row, one task is found on the same column a $x$.
So, amount of data sent by the task $x$ to the supernode $b$ is $(64/(PQ))$. 
Since there are $128$ tasks present in supernode $a$,
the amount of data sent from $a$ to $b$ is $(128\times 64/(PQ))$. Since $PQ=128\NS$,
we see that the amount of data is $64/\NS$. All this data is striped on the $\ND$ D-links
going from $a$ to $b$. Hence, the load on the D-link $e$ is $64/\NS\ND$.
Therefore, the D-link throughput of the row-wise mapping $\pi$ is 
\begin{eqnarray}
\label{eqn:GGG}
\tau_{\D}(\pi) = \frac{\NS\ND}{64} \times 4\times 10 = 20\Hf,
\end{eqnarray}
(since $\Hf=\NS\ND/32$).
A similar calculation shows that the D-link throughput of the column-wise mapping is also $20\Hf$.

A comparison of the Halo and Transpose pattern is noteworthy.
In Transpose, any task $x$ sends $(1/2)$ units of data to the tasks in its row,
all of which constitute intra-supernode communication;
the reaming $(1/2)$ units is sent to tasks along the column of $x$.
Thus, at most $(1/2)$ units data is sent to other supernodes.
The tasks found on the column of $x$ are equally distributed across the supernodes.
As a result, each task will send an equal amount of data every other supernode.
So, the load on the D-links is automatically balanced. 
Hence, in contrast to Halo, we do not need sophisticated techniques (such as mod-coloring) for obtaining 
load balancing in the case of Transpose.

Both row-wise and column-wise strategies are equally good with respect to D-link throughput.
However, their L-link throughput may be different although this happens only in certain 
boundary conditions related to the grid sizes.
This motivates the design of a hybrid scheme where we choose either row-mapping or column-mapping depending
on the grid size. The next step is to map tasks to individual nodes within supernodes.
For this we adopt a simple sequential mapping strategy.
It can be proved that the LR-link throughput is then at least $80$ {\GBs} and the LL-link throughput is
at least $134$ {\GBs} with the actual values depending on the grid size.
The overall throughput is then given by:
\begin{equation}
\label{eqn:HHH}
\min\{20\Hf, 80\} \GBs.
\end{equation}

\par\noindent
Boundary cases and derivations are deferred to the Appendix. 

\section{Mapping Strategies for Indirect Routing}
\label{indirect}
Indirect routing uses an intermediary supernode different from the source and
destination supernodes as a ``bounce'' point. 
We describe indirect routing in  detail and study 
the mapping problem for the Halo and Transpose patterns.

\subsection{Routing Scheme}
\label{indirect-routing}
Inter-supernode communication 
uses an intermediate supernode as a bounce point that redirects the communication to the 
destination supernode. There are a total of ${\NS}\times {\ND}$ D-links from the source supernode.
Each such D-link defines a path as follows.
Pick any D-link $e$ that connects the source supernode to an intermediate supernode $x$. 
Consider the bucket on the intermediate supernode where D-link $e$ terminates.
The path is completed by using the unique D-link originating from the same bucket to
travel to the destination supernode.

\begin{figure}
\begin{center}
\includegraphics[width=3.0in]{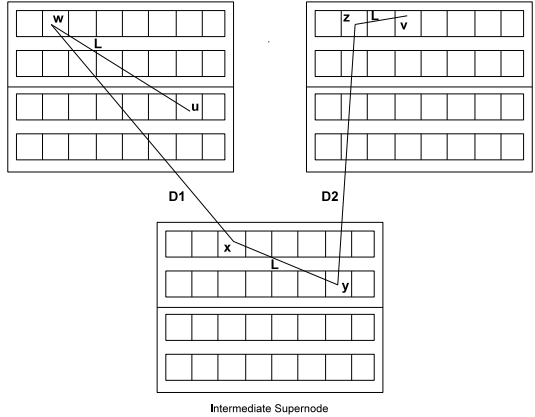}
\end{center}
\caption{Indirect routing}
\label{fig:DDD}
\end{figure}

For an illustration, refer to Figure \ref{fig:DDD} that shows one of the ${\NS}\ND$ indirect paths.
This path is defined by the D-link $D_1$ from supernode $a$ to some intermediate supernode $c$. 
Link $D_1$ originates on node $w$ and terminates on node $x$ in supernode $c$. 
Since $w$ is in bucket 0, $x$ is also in bucket 0. The system wiring dictates that there be a D-link 
originating from bucket 0
of supernode $c$ going to supernode $b$; in the illustration, this link originates from node $y$
and lands on node $z$ in supernode $b$. Thus, the indirect path defined by $D_1$ is 
$w \rightarrow x \rightarrow y \rightarrow z$. 
Notice that one could have used the D-link originating from bucket 1 (instead of bucket 0)
of intermediate supernode $c$ to reach supernode $b$. However, such a path is not
supported by the routing software used in the system.
For the second D-hop, the path is constrained to use the unique D-link originating from the same bucket where 
the first D-hop $D_1$ lands.

For intra-supernode communication we use the same striped routing method as in Direct
routing (See Section~\ref{direct-routing}).

\subsection{Principles}
\label{principles-indirect}
We now discuss some general principles that are useful 
in designing good mapping strategies with indirect routing.

A crucial difference between direct and indirect routing schemes is that
the latter tends to  balance load on the D-links.
Consider a node $u$ sending data to node $v$ in a different 
supernode. In the case of direct routing, this data will be sent over
the bundle of $\ND$ D-links connecting the two supernodes.
In the case of indirect routing, the data will be striped
over {\em all} the $\NS\ND$ D-links going out of $a$.
As a result the load is well balanced on the D-links making 
it easier to design good indirect routing mapping
strategies.

Let $\delta_{\max}$ be the maximum amount of data
sent from any supernode to any other supernode (where the maximum is taken
over all pairs of supernodes).
Then, the D-link throughput will be $\delta_{\max}/(\NS\ND)$.
In order to obtain high D-link throughput under indirect routing,
it is sufficient to simply minimize the load out of a supernode, or
$\delta_{\max}$.

Now let us consider the case of L-links.
Consider an L-link in supernode $a$ from node $u$ to node $v$.
As before, consider inter-supernode communication using paths of the form L-D-L-D-L.
Link $e$ will be used as the first L-hop whenever $u$ sends data to any other supernode
and the data will be striped uniformly over the 32 L-links originating from $u$.
Similarly, link $e$ will be used as the last L-hop whenever $u$ receives data from 
any other supernode and the data will be striped uniformly over the 32 L-links terminating at $u$.
So, the load on $e$ due to the first and last L-hops are determined purely
by the sum of the amount of sent by $u$ and received by $v$ to/from other supernodes.
A good mapping strategy should minimize this sum over all the L-links.
Thus, the main consideration is to  reduce the amount of data sent by individual nodes to
other supernodes.
However, the middle L-hop poses an interesting issue which we discuss in the
context of Halo pattern.

\subsection{Mapping Strategies for the Halo Pattern}
\label{halo-indirect}
In this section, we discuss mapping strategies for the Halo communication pattern, under indirect routing.
We also show that indirect routing is better than direct routing for the Halo pattern.
Bhatale et. al.~\cite{SCpaper} applied the blocking schemes for the Halo pattern (c.f.
Section \ref{halo-direct})
also under the indirect routing scheme. 
We shall argue that the supernode blocking scheme under random mapping is a good strategy in this scenario.

As our principles (Section~\ref{principles-indirect}) suggest, 
it is sufficient for the mapping strategy to reduce the D-link loads under indirect
routing scheme.
In Section \ref{halo-direct}, we also observed that blocking is beneficial for 
load reduction for Halo.

Consider a Halo job grid of size $P\times Q$, where $PQ=n=128\NS$
is the number of tasks. Let us fix a block size of $\alpha\times \beta$.
The data sent out of any supernode is $64(\alpha+\beta)/(\alpha\beta)$ (see Equation \ref{eqn:CCC}).
We wish to ensure that $\alpha\beta$ divides $128$ so that each supernode
gets an integral number of blocks. Under this condition the block size
minimizing the amount of data is $8\times 16$. 
So, we see that the block size of $\alpha=8$ and $\beta=16$ is the best choice (i.e., supernode blocking).
We see that each supernode will get exactly one block and there will be $\NS$ blocks.
In this case, the data sent of a supernode is $12$ units.
It can be shown that the D-link throughput is:
\begin{eqnarray}
\label{eqn:DDD}
\tau_{\D}(\pi) = (\NS\ND/24)\times 4\times 10 = 5(\NS\ND)/3 \GBs.
\end{eqnarray}

Now let us consider the case of LR and LL links.
Given our block size of $8\times 16$, each supernode gets one block.
The L-link load is determined by how the tasks within a block
are mapped to processors within the supernode.
Here, a simple strategy that 
divides the block into $2\times 2$ {\em quads}
and maps each quad to a node of the supernode in a sequential manner suffices.
It can be shown that intra-supernode communication 
does not become the bottleneck due to striping.
Consider inter-supernode communication, where the paths are of the form L-D-L-D-L.
Let $e$ be a link going from a node $u$ to a node $v$ in some supernode.
We can show that both the amount of data sent by $u$ to the supernodes incident on $v$
and the amount of data received by $v$ from the supernodes incident on $u$
are minimized. As suggested by our principles (Section \ref{principles-indirect}),
the above load will be well distributed over the L-links.
However, the middle L-hop poses an interesting issue;
this is discussed in the Appendix. 

From our analysis it is clear that with respect to the D-link throughput,
the indirect routing scheme outperforms the direct routing scheme.
Our experimental result show that it is indeed the case for the overall throughput as well.

\subsection{Mapping Strategies for Transpose Pattern}
\label{transpose-indirect}
In this section, we present a brief overview of mapping strategies for Transpose.
We shall argue that, in contrast to the Halo pattern, direct routing is better than indirect routing
for the Transpose pattern.

The row-mapping and column-mapping strategies (Section \ref{transpose-direct}) are
also good under 
indirect routing 
because they provide good load reduction. It can be shown that the D-link throughput is $10\Hf$ \GBs.
We saw that the same mapping offers a D-link throughput of $20\Hf$ \GBs.
The reason for the reduction in the throughput in the current scenario is that indirect routing uses
two D-hops, whereas direct routing uses only one D-hop.

The hybrid mapping approach we discuss in Section~\ref{transpose-direct} applies
also to the case of indirect routing.
The overall throughput can be shown to be:
\begin{equation}
\label{eqn:JJJ}
\tau(\pi) \geq \min\{10\Hf, 320/(4+\ND)\}.
\end{equation}
A formal derivation is presented in the Appendix.

From Equations (\ref{eqn:HHH}) and (\ref{eqn:JJJ}), we see that 
direct routing outperforms indirect routing with grid sizes 
where the smaller dimension is less than $128$.

\section{Experimental Study}
\label{experiment}
In order to evaluate the different mapping schemes presented in this paper, 
we have developed a simulator that takes as input:
(i) the system configuration ($\NS$ and $\ND$); (ii) a job pattern (iii) a mapping; (iv) the routing scheme. 
The simulator performs all the communications according to the input parameters and 
calculates the load on every link. Then, it computes the maximum load on LL, LR and D links separately.
Using these maximum loads, it computes the overall throughput of the mapping scheme.

Using the simulator, we studied the efficacy of the 
various mapping schemes for different job patterns.
The following parameters were considered:
(i) Number of supernodes $\NS$ (varied from $16$ to $512$ in powers of two);
(ii) Number of D-links $\ND$ (varied from $1$ to $16$ in powers of two);
(iii) Routing schemes: Direct and indirect.

\subsection{Halo Pattern}
For the Halo pattern, we studied the following eight routing schemes: 
(i) Default (DEF); (ii) Node sequential (NODE SEQ); (iii) Node random (NODE RND); (iv) Drawer Sequential (DRW SEQ); 
(v) Drawer random (DRW RND); (vi) Supernode sequential (SN SEQ); (vii) Supernode random (SN RND); 
(viii) Mod-coloring (MOD CLR).
Though node-sequential and the node-random schemes were better than the default scheme,
they performed poorly in comparison to other schemes.
Consequently, in the interest of space and readability, we have omitted the results for these two schemes.

{\bf {\em n}$_s$ = 32, {\em n}$_d$ = varying, Routing = Direct: }
The results for direct routing are shown in Table~\ref{halo-ns-fixed-direct}.
In most of the cases, D-links were the bottleneck. Occasionally, the LR-links
became the bottleneck - these cases are marked with (*) in the table.

The experimental results show that a default mapping fairs poorly. 
With regards to supernode-blocking and the drawer-blocking, recall that our earlier analysis 
shows that the supernode-blocking scheme achieves better load reduction on the D-links
in comparison to the drawer-blocking; whereas, drawer-blocking achieves better load distribution.
The results indicate that two supernode-blocking schemes are better than the drawer-blocking schemes.
The mod-coloring scheme outperforms all the other schemes. Our earlier analysis had shown that 
mod-coloring achieves a good balance between load reduction and load distribution.
For the cases of $\ND=1,2$, the D-link is the bottleneck; 
hence, a throughput of $20\ND$ {\GBs} is achieved, as our analysis had indicated.
This is twice that of the supernode-blocking schemes.
For $\ND\geq 4$, even though the D-link throughput is twice that of the other schemes,
the overall throughput is diminished as the LR-links become the bottleneck.
In terms of overall throughput, the mod-coloring scheme outperforms other schemes by as much as factor of 2.

\begin{table}
\begin{center}
\begin{footnotesize}
\begin{tabular}{|c||c|c|c|c|c|c|}
\hline
${\ND}$ & DEF & DRW & DRW & SN & SN & MOD\\
        &     & SEQ & RND & SEQ & RND & CLR\\
\hline
\hline
1&     2&       5&       8&      10&      10&      20\\
2&     5&       10&      16&      20&      20&      40\\
4&    10&      20&      33&      40&      40&      64 (*)\\
8&    20&      40&      66&      80&      80&     107 (*)\\
16&   40&      80&     120 (*)&     160&     128 (*)&     160 (*)\\
\hline
\end{tabular}
\end{footnotesize}
\end{center}
\caption{\scriptsize Job=Halo; $\NS=32$; $\ND$=varying; Routing=Direct}
\label{halo-ns-fixed-direct}
\end{table}

{\bf {\em n}$_s$ = varying, {\em n}$_d$ = 4, Routing = Direct: }
The results under direct routing are shown in Table~\ref{halo-nd-fixed-direct},
for the fixed value of $\ND=4$ and varying $\NS$.
Consider the mod-coloring scheme.
In this case the guaranteed D-link throughput is $20\ND=80$ \GBs.
In conjunction with Table~\ref{halo-ns-fixed-direct},
it seems that for $\ND\geq 4$, LR becomes the bottleneck (irrespective of $\NS$).
Despite this phenomenon, the mod-coloring scheme outperforms the other schemes.
A careful study of the LR-throughput behavior of the mod-coloring scheme
would be interesting and may lead to a better mapping scheme.

\begin{table}
\begin{center}
\begin{footnotesize}
\begin{tabular}{|c||c|c|c|c|c|c|}
\hline
${\NS}$ & DEF & DRW & DRW & SN & SN & MOD\\
        &     & SEQ & RND & SEQ & RND & CLR\\
\hline
\hline
16  &   10   &   20   &   29   &   40   &   40   &   64 (*)\\
32  &   10   &   20   &   33   &   40   &   40   &   64 (*)\\
64  &    5   &   20   &   37   &   40   &   40   &   64 (*)\\
128 &    5   &   10   &   38   &   40   &   40   &   64 (*)\\
\hline
\end{tabular}
\end{footnotesize}
\end{center}
\caption{\scriptsize Job=Halo; $\NS$=varying; $\ND$=4; Routing=Direct}
\label{halo-nd-fixed-direct}
\end{table}

{\bf {\em n}$_s$ = 32, {\em n}$_d$ = varying, Routing = Indirect: }
The results under indirect routing are shown in Table~\ref{halo-ns-fixed-indirect} with 
the type of bottleneck specified next to the throughput figure.
In our analysis, we had observed that the D-link throughput is directly proportional to $\ND$
and is determined by the block size.
For the case of $\ND=1$, the supernode blocking schemes use the optimum block size
outperforming the other blocking schemes.
As $\ND$ gets larger, the D-link throughput of all the schemes increases
and bottleneck shifts from the D-links to the other links.
When $\ND=2$, mostly the LR becomes the bottleneck
and for $\ND\geq 4$, typically the LL becomes the bottleneck.
We performed additional experiment to better understand this phenomenon.
From our discussion regarding the middle L-hop, we predicted that the middle L-hop could be the 
cause of the bottleneck. In order to verify the prediction,
we disabled the accounting for the load due to the middle L-hop in our program.
We found that bottleneck shifted to a different type of link
and concluded that the middle L-hop is indeed a reason for the bottleneck on the L-links.
Notice that when $\ND\geq 4$, the LR-links are not used as middle L-hop
and hence, LL-links become the bottleneck.
These results suggest that a deeper study of the middle L-hop issue is required.

\begin{table}
\begin{center}
\begin{footnotesize}
\begin{tabular}{|c||c|c|c|c|c|}
\hline
${\ND}$ & DEF & DRW & DRW & SN & SN \\
        &     & SEQ & RND & SEQ & RND \\
\hline
\hline
1  &  20 (D)  &     36  (D) &   27  (D)&    53  (D)&    53 (D)\\
2  &  34 (LR)  &     58  (LR) &   53  (D)&    91  (LR)&    96 (LR)\\
4  &  80 (D)  &    128  (LL) &  107  (D)&   134  (LL)&   174 (LR)\\
8  & 103 (LL)  &     93  (LL) &  127  (LL)&   183  (LR)&   167 (LL)\\
16 &  64 (LL)  &    179  (LL) &  103  (LL)&   168  (LL)&   148 (LL)\\
\hline
\end{tabular}
\end{footnotesize}
\end{center}
\caption{\scriptsize Job=Halo; $\NS$=32; $\ND$=varying; Routing=Indirect}
\label{halo-ns-fixed-indirect}
\end{table}


\subsection{Transpose Pattern}
{\bf {\em n}$_s$ = 32, {\em n}$_d$ = varying, Routing = Direct: }
The results under direct routing are shown in Table~\ref{transpose-ns-fixed-direct}, with
the type of bottleneck specified next to the throughput figure.
We experimented with our hybrid scheme and compared it against the supernode blocking scheme.
These results indicate that row-wise/column-wise mapping is superior to the blocking schemes.
The table also conforms with the throughput analysis predicted by Equation \ref{eqn:HHH}.
\begin{table}
\begin{center}
\begin{footnotesize}
\begin{tabular}{|c||c|c|}
\hline
${\ND}$ & SN  & Hybrid\\
        & SEQ & Scheme\\
\hline
\hline
1   & 2 (D)  &  20 (D)\\
2   & 5 (D)  &  40 (D)\\
4   & 10 (D) &  80 (D)\\
8   & 20 (D) &  80 (LR)\\
16  & 40 (D) &  80 (LR)\\
\hline
\end{tabular}
\end{footnotesize}
\end{center}
\caption{\scriptsize Transpose; $\NS$=32; $\ND$=varying; Routing=Direct}
\label{transpose-ns-fixed-direct}
\end{table}

\bibliographystyle{IEEEtran}
\bibliography{main}

\appendix
\section{Mod Coloring : Proof of Lemma {\ref{LEM:FFF}}}
\label{app:mod-color}
Let $p\times q$ be the input grid.
Our goal is to obtain a perfect coloring of the grid.
The notion of nice permutations is useful for this purpose.

Let $\Omega=\{a_1, a_2, \ldots, a_q\}$ be a set of $q$ elements and 
$\sigma_1$ and $\sigma_2$ be two permutations over $\Omega$.
For $0\leq i<q$, let $\sigma_1(i)$ and $\sigma_2(i)$ represent the element in the $i$th
position under $\sigma_1$ and $\sigma_2$, respectively.
We view the two permutations as a $2\times q$ matrix by taking $\sigma_1$ to be the first row
and $\sigma_2$ to the second row. 
Consider any element $a$. 
Each element $a$ appears twice in the matrix, once in each row.
Let $l_1, r_1$ and $d$ represent
the elements found in positions left, right and below the copy of $a$ in the first row
(for the two corner positions, left and right are obtained via wrap-arounds).
Similarly, let $l_2, r_2$ and $u$ represent the elements appearing in positions left, right and above
the copy of $a$ in the second row. We call the above six elements as the {\em neighbors} of $a$.
The elements $u$ and $d$ are called the up and down neighbors of $a$.
The permutations $\sigma_1$ and $\sigma_2$ are said to be a {\em nice pair},
if for all elements $a\in \Omega$, all the six neighbors of $a$ are distinct.
An example is shown in Figure \ref{fig:EEE} for the set $\Omega=\{0,1,2,\ldots,7\}$.
The neighbors of the element $5$ are $l_1=4$, $r_1=6$, $l_2=0$ and $r_2=2$;
the up-neighbor is $u=7$ and the down-neighbor is $d=3$.
We see that all these six neighbors are distinct. It can be verified that
the two permutation form a nice pair. The next lemma shows how to construct a nice 
pair of permutations, when $q\geq 8$ is a power of two.

\begin{figure}
\begin{center}
\begin{tabular}{|c||c|c|c|c|c|c|c|c|}
\hline
$\sigma_1$ & 0 & 1 & 2 & 3 & 4 & 5 & 6 & 7\\
\hline
$\sigma_2$ & 2 & 7 & 4 & 1 & 6 & 3 & 0 & 5\\
\hline
\end{tabular}
\end{center}
\caption{An example nice pair of permutations}
\label{fig:EEE}
\end{figure}

\begin{lemma}
\label{lem:BBB}
Let $\Omega$ be any set of $q$ elements, where $q$ is a power of two and $q\geq 8$.
Then, there exists a nice pair of permutations for $\Omega$.
\end{lemma}

The lemma is proved in the next section.
For now, we assume the lemma and complete the proof of Lemma \ref{LEM:FFF}.

We index the rows of the $p\times q$ grid from $0$ to $p-1$ and the columns from $0$ to $q-1$.
We use $\pair{x}{y}$ to mean the grid on the $x$th row and $y$th column.

\begin{figure}
\begin{center}
\includegraphics[width=2.0in]{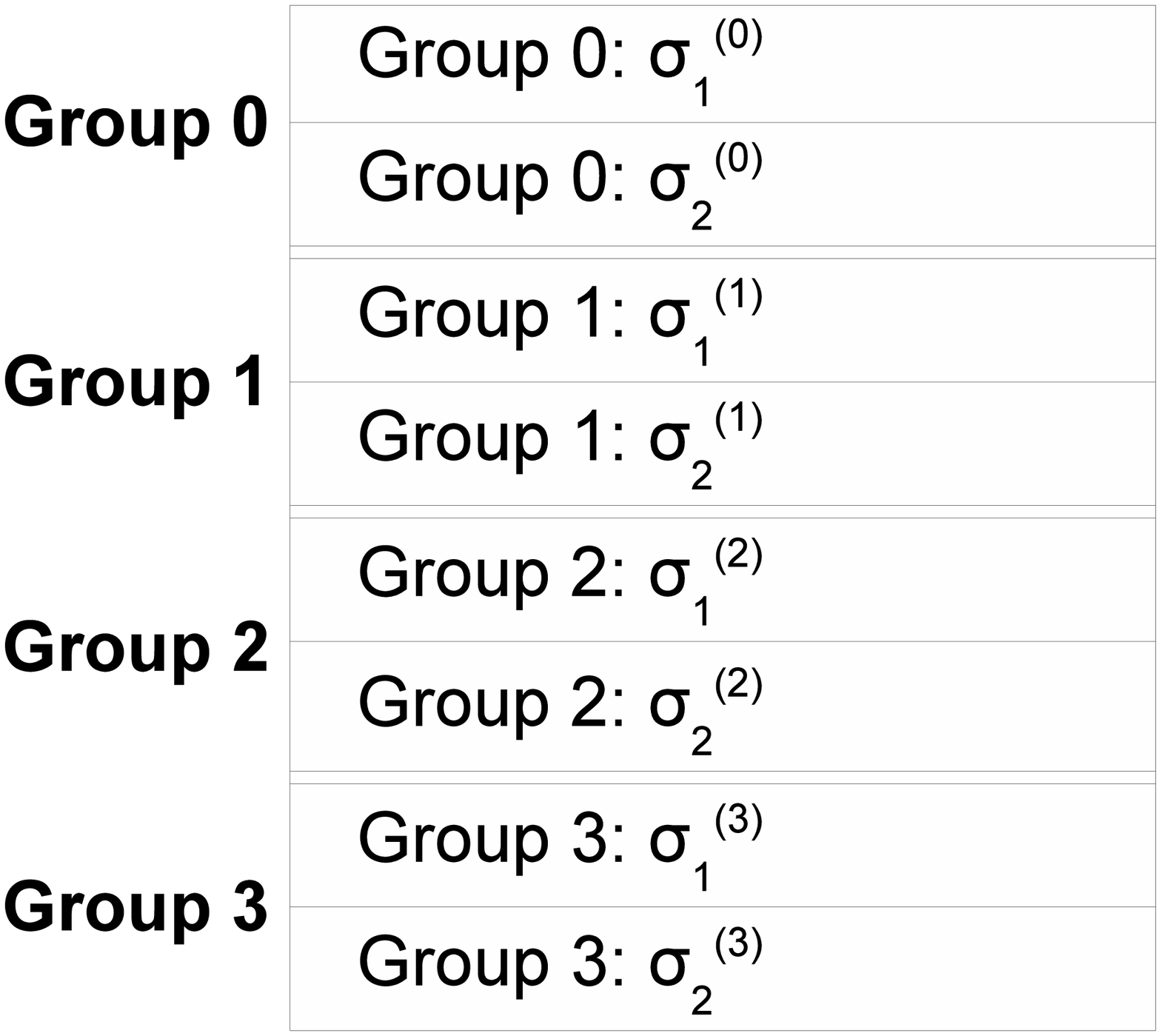}
\end{center}
\caption{Illustration for Lemma \ref{LEM:FFF}}
\label{fig:MMM}
\end{figure}

Divide the set of colors in to $(p/2)$ groups each containing $q$ colors in a sequential manner as follows: 
for $0\leq j<(p/2)$, the group $j$ is defined to be $\{jq,jq+1, \ldots, (j+1)q-1\}$.
For $0\leq j<(p/2)$, apply Lemma \ref{lem:BBB} on the group $j$ and obtain a pair of permutations
$\sigmap{j}{1}$ and $\sigmap{j}{2}$. Divide the rows of the grid into $(p/2)$ groups each
consisting of two rows each in a sequential manner (so the $j$the group will have rows $2j$ and $2j+1$).
For $0\leq j<(p/2)$ the first row of the $j$th group is colored using the permutation $\sigmap{j}{1}$ and
the second row is colored using the permutation $\sigmap{j}{2}$.
Formally, for $0\leq x<p$ and $0\leq y<q$, the grid point $\pair{x}{y}$ will be colored as follows.
Let $j=\lfloor x/2 \rfloor$;
if $x$ is an even number then assign the color $\sigmap{j}{1}(y)$; 
whereas, if $x$ is an odd number then assign the color $\sigmap{j}{2}(y)$.
See Figure \ref{fig:MMM} for an illustration; here $p=8$.

We now argue that the above coloring is perfect. 
First consider the case where $p\geq 6$.
Let $c$ be any color and let $j$ be the group to which it belongs.
The color $c$ appears in the once each in the rows $2j$ and $2j+1$.
Let $l_1, r_1, u_1$ and $d_1$ be the colors appearing in the left, right, up and down positions
of $c$ appearing the first row, respectively. 
Similarly, let $l_1, r_2, u_2$ be the colors in the neighboring positions of $c$ appearing in the second row.
Of the eight neighboring colors, six of them ($l_1, r_1, d_1, l_2, r_2$ and $u_2$) belong
to the same color group as $c$. By the niceness property ensured by Lemma \ref{lem:BBB},
all these six colors are distinct. The color $u_1$ belongs to the color group given by $(j-1)$ (modulo $p/2$).
similarly, the color $d_2$ belongs to the color group given by $(j+1)$ (modulo $p/2$).
Since $p\geq 6$, $j$, $(j-1)$ and $(j+1)$ are all different color groups.
This shows that $u_1$ and $d_1$ are different, and they are also distinct from the above six other neighbors.

\begin{figure}
\begin{center}
\includegraphics[width=2.0in]{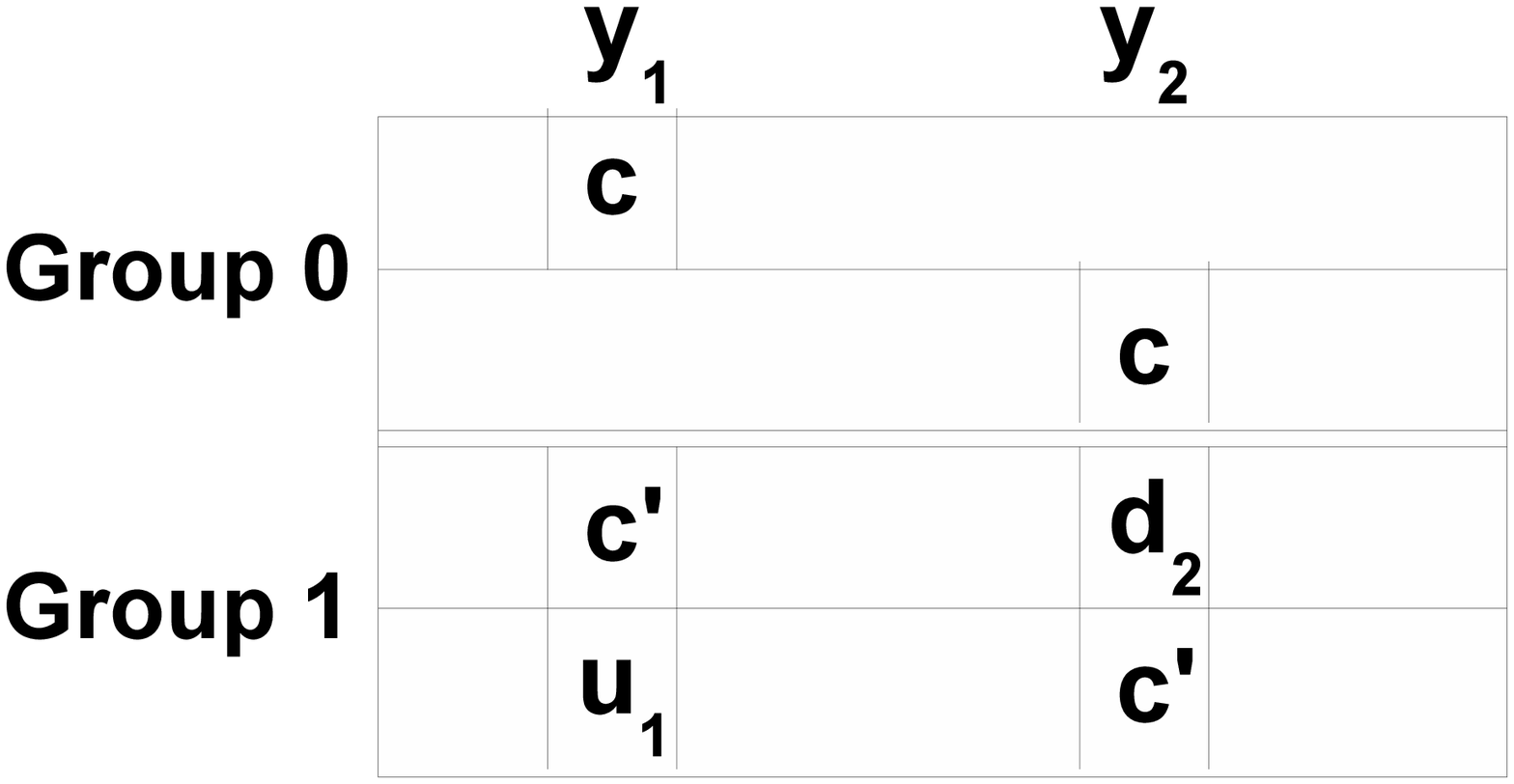}
\end{center}
\caption{Illustration for Lemma \ref{LEM:FFF}}
\label{fig:NNN}
\end{figure}

Let us consider the case of $p=4$. 
The issue here is that the number of groups is only two and so, for any color $c$,
the $u_1$ and $d_2$ neighbors will belong to the same group.
Nevertheless, we argue below that these two are distinct.
Going through the proof of Lemma \ref{lem:BBB}, we see that it applies the
same construction on any given set of elements $q$ elements $\Omega$.
Therefore, for any color $c$ belonging to group $0$ and $c'$ belonging to group $1$, 
if $c$ and $c'$ appear in the same positions under 
first set of permutations ($\sigmap{0}{1}$ and $\sigmap{1}{1}$) then they also appear in the
same position under the second set of permutations ($\sigmap{0}{2}$ and $\sigmap{1}{2}$).
Based on the property, we now argue that the coloring scheme is perfect.
Refer to Figure \ref{fig:NNN} for an illustration. 
Let $c$ be any color from the group $0$. Consider the $u_1$ and $d_2$ neighbors of $c$.
Let $y_1$ and $y_2$ be the positions in which $c$ appears in the permutations $\sigmap{0}{1}$
and $\sigmap{1}{2}$, respectively. Then, $u_1=\sigmap{1}{2}(y_1)$ and $d_2=\sigmap{1}{1}(y_2)$.
Let $c'=\sigmap{1}{1}(y_1)$.
By the property stated above, $\sigmap{1}{2}(y_2)$ is also $c'$.
Notice that $u_1$ and $d_2$ are the up and down neighbors of $c'$.
By the properties of nice permutations, these are distinct.
A similar argument applies for the case of colors in group 1.
\qed

\subsubsection*{Nice Permutations (Proof of Lemma \ref{lem:BBB})}
Without loss of generality, assume that $\Omega = \{0, 1, 2, \ldots, q-1\}$.
We define $\sigma_1$ to be the identity permutation: for $0\leq i< q$, $\sigma(i)=i$.
The permutation $\sigma_2$ is defined as follows: for $0\leq i<q$, 
$\sigma_2(i)=5i+2 \pmod{q}$. It is easy to see that $\sigma_2$ is indeed a permutation
(because $5$ is a prime number and so, it is a generator for $Z_q$ -- the additive group modulo $q$).
We proceed argue that $\sigma_1$ and $\sigma_2$ form a nice pair.
Throughout the discussion below all calculations are performed modulo $q$.

Consider any element $0\leq a<q$. Five of its neighbors are:
\begin{center}
\begin{tabular}{ccc}
$l_1 = a-1$ & $r_1 = a+1$ & $d=5a+2$\\
$l_2 = a-5$ & $r_2 = a+5$ &
\end{tabular}
\end{center}
where all the above calculations are performed modulo $q$. We will consider the sixth up-neighbor $u$ later.
It is easy to see that $l_1, r_1, l_2$ and $r_2$ are all distinct 
(because $q\geq 8$ and so, $-1, +1, -5$ and $+5$ are all distinct modulo $q$).
We next argue that $5a+2$ is distinct from the other four numbers.
For an element $0\leq i\leq q$, let $g(i)$ be the set generated by $i$,
i.e., $g(i)=\{ix\pmod{q}~:~x\geq 1\}$.
Now, by contradiction, suppose, $5a+2=a+\delta$, for some $\delta\in \{-1, +1, -5, +5\}$.
This implies that $4a+2=\delta$. Notice that $4a+2\in g(2)$, but $\delta\not\in g(2)$,
(because $q$ is an even number and $q\geq 8$).
This shows that $5a+2$ is distinct from the other four numbers. 

We next prove that the up-neighbor $u$ is different from the other five neighbors.
Let us first consider the case of $l_1$.
Notice that the down-neighbor of $l_1$ is $5(a-1)+2$ and $5(a-1)+2\neq a$ 
(because $q$ is an even number and so, $3\not\in g(4)$).
This means that the down-neighbor of $l_1$ is not $a$.
On the other hand, the down-neighbor of $u$ is $a$. Therefore, we get that $l_1\neq u$.
A similar argument shows that $r_1\neq u$.
Let us now consider the case of $l_2$. Notice that $a=\sigma_2(u)$ and so,
$l_2=5u-3$. It is easy to see that $5u-3\neq u$ (because $3\not\in g(4)$).
This shows that $l_2\neq u$. A similar argument shows that $r_2\neq u$.
Finally, let us show that $d\neq u$. To prove the claim, it suffices to show that
the down-neighbor of $d$ is not $a$. The down-neighbor $d$ is $5d+2=5(5a+2)+2=25a+12$.
We see that $25a+12\neq a$ (because $q$ is a power of two and so, $-12\not\in g(24)$).
\qed

\section{Middle L-hop Issue} 
As mentioned in Section \ref{halo-indirect} regarding Halo pattern under indirect routitng, 
the middle L-hop poses an interested issue. This is discussed here. 

The notion of congruence classes is useful in determining the load contributed by the middle L-hop.
We say that two supernodes are {\em congruent}, if they are incident on the same set of nodes.
Thus, the supernodes are portioned into as many congruence classes as there are nodes in a
bucket.
For example, in Figure \ref{fig:AAA}(b), there are 16 congruence classes each containing two supernodes;
supernodes $\{0,16\}$ belong to one congruence class, and $\{1,17\}$ belong to another.
Two L-links going from nodes $u_1$ to $v_1$
and $u_2$ to $v_2$ are said to be congruent, if the same set of supernodes are incident on $u_1$ and $u_2$
{\em and} the same set of supernodes are incident on $v_1$ and $v_2$.
For instance, in Figure \ref{fig:AAA}(b), the L-links going from node $0$ to $8$, and the L-links going from
node $16$ to $24$ in all the supernodes form a congruence class.
The number of L-links in a congruence class is $\NS\ND$.

{\bf Middle L-hop: }
At a first glance, it may seem that how the $\NS$ blocks are mapped to the $\NS$ supernodes
is not crucial. However, we argue that the above mapping is important and that it
determines
how the middle L-hop load is distributed over the L-links.
Consider an L-link $e$ in supernode $a$ from node $u$ to a node $v$.
There is a set of $\Hf$ supernodes incident on the node $u$ and similarly, there is a set of $\Hf$
supernodes incident on the node $v$. 
The link $e$ will be used as the middle L-hop whenever any data is sent from the former set of supernodes
to the latter set of supernodes. This data will be striped over the $\NS\ND$ L-links in
the congruence class
of $e$.
Let us analyze the case where this amount of data will be large.
A block mapped to a supernode has four neighboring blocks mapped to four other supernodes;
thus, each supernode can be viewed as having four neighboring supernodes.
Now, unless the block-to-supernode mapping is done carefully,
the set of supernodes incident on $u$ may find many of their neighboring supernodes
being incident on the node $v$ causing link $e$ to receive a lot of data.
To avoid the middle L-hop issue, the block-to-supernode mapping must ensure that
the neighbors of any congruence class of supernodes are spread across the other congruency classes.
A simple way to achieve the above goal is to map the blocks to supernodes in a random fashion. 
Notice that this method is the same as random supernode blocking 
scheme proposed by Bhatale et al.~\cite{SCpaper} (see Section~\ref{halo-direct}).
More sophisticated techniques, for instance based on mod-coloring,
may provide better mapping schemes. This is left as future work.

\section{Transpose Pattern}
\subsection{Transpose Pattern : Direct Routing}
Here we discuss the boundary cases mentioned in Section \ref{transpose-direct} and also present
the derivations for the LL and LR link throughput for Transpose under direct routing.

We discussed that with respect to D-links throughput, both row-wise and column-wise
mappings are equally good. 
However, two minor issues dealing with boundary cases distinguishes the two schemes.
The first issue is regarding the LR-link throughput.
Consider the case of row-wise mapping. Since we assume that $P$ is a power of two,
each row will fit within a drawer (if $P\leq 32$), or it will span two drawers (if $P=64$)
or it will span the entire supernode (if $P=128$). In the first case, there will be
no load on the LR-link due to intra-supernode communication.
In the second case, the load on any LR-link (due to intra-supernode communication) will be $(1/8)$ units;
on the other hand, in the third case, the load will be $(1/16)$.
This is because, in the second case only LR links connecting two pairs of drawers are used;
whereas in the third case, LR links across all the four drawers are used.
Hence, for the case where $P=64$ and $Q\neq 64$,  the column-wise mapping will be better;
similarly, for the case where $P\neq 64$ and $Q=64$, the row-wise mapping will be better.
The second issue is regarding another boundary case.
Both the schemes require that either $P$ or $Q$ is at most $128$.
Thus, for the grid of size $128\times 256$, we must choose column-wise mapping
and similarly, for the grid of size $256\times 128$, we must choose row-wise mapping.
Based on the above discussions, it is easy to design a simple hybrid scheme 
which choose between row-wise and column-wise mapping based on the input grid dimensions.

Let us compute the LR-link throughput of the above hybrid scheme.
Consider an LR-link $e$ going from a node $u$ to a node $v$ in some supernode $a$.
Let us first focus on inter-supernode communication, which uses paths of the form L-D-L.
Node $u$ has four processors each sending approximately $(1/2)$ units of data to its 
neighbors uniformly distributed across the other supernodes.
Thus, a total load of $2$ units is distributed over the $32$ L-links originating from $u$.
So, the load on $e$ is approximately $1/16$ units while being used as the first L-hop.
Similarly, node $v$ receives $2$ units of data over its $32$ L-links.
Hence, the load on $e$ is approximately $1/16$ units while being used as the second L-hop.
So, the load on the link $e$ is $(1/8)$. 
We already saw that the load on any LR-link due to intra-supernode communication is at most $(1/8)$.
So, the total load on any LR-link $e$ is at most $(1/4)$. So, the LR-link throughput is at least $80$ \GBs.

Let us now analyze the case of LL-links. First consider intra-supernode communication.
The load on LL-links will be maximum when each row (or column) fits within a drawer;
in this case, the load can be shown to be $(1/2)$ units.
So, the load on any LL-link due to intra-supernode communication is at most $(1/2)$ units.
The load due to inter-supernode communication is $(1/8)$ (the same as LR).
So, the total  load on any LL-link is at most $5/8$ units.
Therefore, the LL-link throughput is at least $134$ \GBs.


\subsection{Transpose Pattern : Indirect Routing}
In Section \ref{transpose-indirect}, we presented a brief overview of the mapping strategies for
the transpose pattern under indirect routing. A detailed discussion is given below.

We first focus on the D-links. The properties of indirect routing ensure that the load will be inherently 
well-balanced on the D-links.
We already saw that row-wise mapping and column-wise mapping provide good
load reduction on the D-links. 
Let us analyze the D-link throughput for row-wise mapping. 
Let the input grid be of size $P\times Q$.
Consider a D-link $e$ going from a supernode $a$ to a supernode $b$.
The paths are of the form L-D-L-D-L and the link $e$ can be used either as 
the first D-hop or the second D-hop.
As discussed in the context of direct routing (Section \ref{transpose-direct}),
$a$ sends $64/\NS$ units of data to every other supernode
and so, the overall data sent from supernode $a$ is $64(\NS-1)/NS$.
This data is striped over all the $\NS\ND$ D-links going out of $a$.
So, the load on $e$ while being used as the first D-hop is $(64/\NS\ND)(1-1/\NS)$.
The link $e$ is used as the second D-hop whenever $b$ receives data from other supernodes.
Thus, the total load on $e$ is $\ell(e) = (128/\NS\ND) (1-1/\NS)$.
The D-link throughput of the row-mapping $\pi$ is 
\[
\tau_{\D}(\pi) = \frac{\NS}{\NS-1} \times \frac{\NS\ND}{128}\times 4 \times 10 = \frac{\NS}{\NS-1}\times 10\Hf
\approx 10\Hf.
\]

Now let us analyze the load on the L-links. 
First, consider the case of inter-supernode communication.
Let $e$ be an L-link going from a node $u$ to a node $v$ in some supernode $a$.
The paths in indirect routing are of the form L-D-L-D-L.
The four processors in the node $u$ send approximately $(1/2)$ units of data to other supernodes.
All this data is striped over all the D-links going out of the supernode $a$
and hence, the data is uniformly striped over the $32$ L-links going out from $u$.
So, the load on the link $e$ being used as the first hop is approximately $(1/16)$.
Similarly, the link $e$ will be used as the last L-hop, whenever $v$ receives data
from other supernodes. A similar calculation shows that the load on $e$ while being
used as the last hop is approximately $(1/16)$.

Let us now consider the case of middle L-hop.
In contrast to the case of Halo (Section \ref{halo-indirect}), 
our mapping ensures that the middle L-hop is not an issue. 
This is because, in Transpose, the amount of data sent from one supernode is another supernode
is uniform across all the pairs of supernodes. 
Therefore, the amount of data sent between congruence classes of supernodes is uniform. 
As a result, all L-links get the same amount of load while being used as the middle L-hop.
Let us explicitly compute the above quantity for the link $e$ going from $u$ to $v$ 
(in an intermediate supernode $a$).
As we saw earlier, the amount of data sent from any supernode to any other supernode is $64/\NS$.
Therefore, the amount of data sent from the supernodes incident on $u$
to the supernodes incident on $v$ is $\Hf\times \Hf\times 64/\NS$.
All this data is striped over the $\NS\ND$ L-links in the congruence class of $e$.
Therefore, the load on $e$ due to the middle L-hop is:
\[
\frac{\Hf\times \Hf \times 64}{\NS}\times \frac{1}{\NS\ND} = \frac{\ND}{16}.
\]

The total load on any L-link due to inter-supernode communication is $(1/16)+(1/16)+(\ND/16)=(2+\ND)/16$.
We saw that the load on LR and LL links (Section \ref{transpose-direct}) due to intra-supernode
communication is at most $(1/8)$ and $(1/2)$ units, respectively.
Therefore, the total load on LR and LL links is at most $(4+\ND)/16$ and $(10+\ND)/16$, respectively.
The throughput can be calculated automatically:
\begin{eqnarray*}
\tau_{\LR}(\pi) & \geq & \frac{320}{4+\ND}\\
\tau_{\LL}(\pi) & \geq & \frac{1344}{10+\ND}
\end{eqnarray*}

The overall throughput is $\tau(\pi) \geq \min\{10\Hf, 320/(4+\ND)\}$.

\end{document}